\documentclass[aps,prd,twocolumn,superscriptaddress,nofootinbib]{revtex4-2}

\usepackage{amsmath,amssymb}
\usepackage{graphicx}
\usepackage{hyperref}
\usepackage{listings}
\usepackage{xcolor}
\begin{document}

\title{Boltzmann Equation Solver for Thermalization}

\author{Jong-Hyun Yoon}
\affiliation{Department of Physics and Institute of Quantum Systems,
Chungnam National University, Daejeon 34134, Republic of Korea}

\begin{abstract}
We present \textsc{Best} (Boltzmann Equation Solver for Thermalization), a Python framework for solving the momentum-resolved Boltzmann equation for arbitrary $n_{\rm in} \to n_{\rm out}$ scattering processes. The collision integral is evaluated directly in $3(n_{\rm total}-2)$ dimensions using the \textsc{Vegas} adaptive Monte Carlo algorithm with vectorized batch evaluation. Momentum conservation is enforced exactly by expressing one particle's momentum through the constraint, while energy conservation is imposed via a narrow Gaussian representation of the delta function. We identify a subtlety in the construction of the collision integral for processes with unequal initial and final multiplicities ($n_{\rm in} \neq n_{\rm out}$) involving identical particles: the full collision rate requires separate evaluation with the observed momentum pinned to each side of the reaction, weighted by the respective particle multiplicities. Failure to account for this leads to systematic violation of energy conservation. The code supports massive particles with time-dependent masses, Bose-Einstein and Fermi-Dirac quantum statistics, multiple coupled species, cosmological expansion with comoving momenta, and both Euler and Heun time integration. Parallelization is achieved by distributing independent momentum grid points across MPI ranks, yielding near-linear scaling to hundreds of cores. We validate the Monte Carlo results against a semi-analytical $2 \to 2$ collision integral with exact energy conservation, following the phase-space reduction of Ala-Mattinen \textit{et al.} As a demonstration, we study thermalization of a massive scalar field through a $2 \leftrightarrow 3$ number-changing process and show that energy conservation is restored only when all identical-particle contributions are correctly summed. The code is publicly available at \url{https://github.com/best-hep/best}.
\end{abstract}

\maketitle

\section{Introduction}
\label{sec:intro}

The Boltzmann equation governs the evolution of particle phase-space distributions in contexts ranging from dark matter freeze-out and freeze-in~\cite{Kolb:1990vq,Hall:2009bx,Binder:2021bmg} to dark sector thermalization~\cite{Profumo:2025uvx,Arcadi:2019oxh} and baryogenesis~\cite{Kolb:1979qa}. Despite its conceptual simplicity, the collision integral poses a formidable computational challenge: for a $2 \to 2$ process, the nine-dimensional phase-space integral must be evaluated at every momentum grid point and at every time step, unless reduced analytically.

Several strategies have been developed to manage this cost. Moment-based approaches such as the number density equation (nBE) and coupled fluid equations (cBE) reduce the problem to ordinary differential equations for integrated quantities, at the price of assuming a specific functional form for the distribution~\cite{Binder:2021bmg}. Even simpler is the relaxation time approximation, $C[f] \approx -(f - f_{\rm eq})/\tau$, which assumes a known equilibrium form and a single timescale, and is correspondingly less accurate: it tracks only the overall collision rate, not the momentum exchange per collision. The standard thermal averaging formalism~\cite{Gondolo:1990dk} is inapplicable when kinetic equilibrium is not maintained, further motivating a direct phase-space approach. Full phase-space solvers (fBE) discretize the momentum variable and evolve each grid point, but typically rely on analytical reduction of the collision integral to lower dimensions. The \textsc{Drake} code~\cite{Binder:2021bmg} implements all three levels within \textit{Mathematica}, using implicit time integration; for elastic scattering it provides both a full numerical treatment of the collision term and a Fokker--Planck approximation. Ala-Mattinen \textit{et al.}~\cite{Ala-Mattinen:2022nuj} developed a method that reduces the $2 \to 2$ collision integral to two dimensions by analytically performing angular integrations and enforcing energy conservation exactly. Du \textit{et al.}~\cite{Du:2021jcj} solved the phase-space Boltzmann equation for $2 \to 2$ and $1 \to 2$ processes using similar analytical reduction, demonstrating significant deviations from the number-density approach in the transition regime between freeze-in and freeze-out. These techniques were originally developed for neutrino transport in the early universe~\cite{Yueh:1976,Hannestad:1995rs,Dolgov:1997mb}, and have since been adapted to dark matter applications.\footnote{For a complementary Monte Carlo approach based on direct simulation of collisions, see Refs.~\cite{Ovchynnikov:2024rfu,Ovchynnikov:2024xyd,Ihnatenko:2025kew}.}

These approaches have proven highly effective for $2 \to 2$ processes. However, a growing class of cosmological scenarios requires processes beyond $2 \to 2$: cannibal dark matter with $3 \to 2$ number-changing interactions~\cite{Carlson:1992fn,Hochberg:2014dra}, semi-annihilation, dark sector cascades involving $2 \to 3$ or $2 \to 4$ scatterings, and particle production through multi-body final states. For such processes, the analytical reduction that makes $2 \to 2$ tractable is generally unavailable, and the phase-space dimensionality grows rapidly: a $2 \to 3$ process requires integration over $3 \times 3 = 9$ dimensions (after fixing the observed and conserved particles), and $2 \to 4$ over 12 dimensions.

Moreover, for processes with $n_{\rm in} \neq n_{\rm out}$ involving identical particles, a subtlety arises in the construction of the momentum-resolved collision integral, which we organize as a general decomposition valid for arbitrary $n_{\rm in} \to n_{\rm out}$. The standard textbook treatment~\cite{Kolb:1990vq} integrates over the observed particle's momentum to obtain the number density equation, where the identical-particle symmetry factors are absorbed into $|M|^2$. In the momentum-resolved case, however, the observed momentum is \emph{fixed}, and contributions from each side of the reaction are structurally distinct integrals that must be summed with the appropriate multiplicities. For symmetric processes ($n_{\rm in} = n_{\rm out}$) with identical particles on both sides, these contributions are equal, but for asymmetric processes they differ, and omitting any of them leads to systematic violation of energy conservation. We discuss this point in detail in Sec.~\ref{sec:identical}.

Adaptive Monte Carlo integration, particularly the \textsc{Vegas} algorithm~\cite{Lepage:1977sw,Lepage:2020tgj}, is well suited to this problem. \textsc{Vegas} uses importance sampling that adapts iteratively to concentrate evaluations where the integrand is large, achieving accuracy that scales as $1/\sqrt{N_{\rm eval}}$ independent of dimension. Combined with vectorized batch evaluation and massively parallel computation over momentum grid points, the direct evaluation of the full collision integral becomes feasible.

In this work we present \textsc{Best}, a Python-based Boltzmann equation solver that implements this approach. The code treats the collision integral for arbitrary $n_{\rm in} \to n_{\rm out}$ processes on equal footing, with the integration dimensionality determined automatically. Key features include support for massive particles with time-dependent masses, Bose-Einstein and Fermi-Dirac quantum statistics with stimulated emission and Pauli blocking, multiple coupled species, cosmological expansion with comoving momenta, and adaptive time stepping with Heun's method. We validate the code against a semi-analytical $2 \to 2$ benchmark and demonstrate correct thermalization for a $2 \leftrightarrow 3$ number-changing process with energy conservation.

\section{Formalism}
\label{sec:formalism}

\subsection{Boltzmann equation}

We consider the Boltzmann equation for the phase-space distribution $f_a(p,t)$ of species $a$ in an expanding universe with scale factor $a(t)$:
\begin{equation}
\label{eq:boltzmann}
\frac{\partial f_a}{\partial t} - Hp\frac{\partial f_a}{\partial p} = \sum_{\rm processes} C_a[f]\,,
\end{equation}
where $H = \dot{a}/a$ is the Hubble rate and the sum runs over all processes involving species $a$. Working with comoving momentum $q = a\,p$, the Liouville term on the left-hand side simplifies to a total time derivative at fixed $q$:
\begin{equation}
\frac{df_a(q,t)}{dt} = \sum_{\rm processes} C_a[f]\,.
\end{equation}
This is the form implemented in the code, with the physical momentum $p = q/a$ used in the collision integral.

\subsection{Collision integral}

For a general process with $n_{\rm in}$ initial-state and $n_{\rm out}$ final-state particles ($n_{\rm total} = n_{\rm in} + n_{\rm out}$), the collision integral for the particle at position $k$ with momentum $\mathbf{p}$ takes the standard form~\cite{Kolb:1990vq}
\begin{equation}
\label{eq:collision_single}
C^{(k)}(\mathbf{p}) = \frac{1}{2E} \int d\Pi \; |{\cal M}|^2 \; \Lambda\,,
\end{equation}
where the Lorentz-invariant phase space measure is
\begin{equation}
d\Pi = (2\pi)^4 \delta^{(4)}\!\!\left(\sum_{i \in \rm in} p_i - \sum_{j \in \rm out} p_j\right) \prod_{l \neq k} \frac{d^3 p_l}{(2\pi)^3 2E_l}\,,
\end{equation}
and the statistical factor is
\begin{equation}
\label{eq:Lambda}
\Lambda = \prod_{i \in \rm in} f_i \prod_{j \in \rm out}(1 \pm f_j) 
- \prod_{j \in \rm out} f_j \prod_{i \in \rm in}(1 \pm f_i)\,,
\end{equation}
where ``in'' and ``out'' denote the initial and final states of the reaction with the upper (lower) sign for bosons (fermions). The index $k$ enters not through $\Lambda$ but through the measure in Eq.~(4): the observed leg is fixed at $\mathbf{p}$ and excluded from the integration, so whether it lies on the $n_\alpha$- or $n_\beta$-side changes which momenta are integrated and hence the value of $C^{(k)}$.

\begin{figure}[t]
\centering
\hspace{3.0em}\includegraphics[width=0.6\columnwidth]{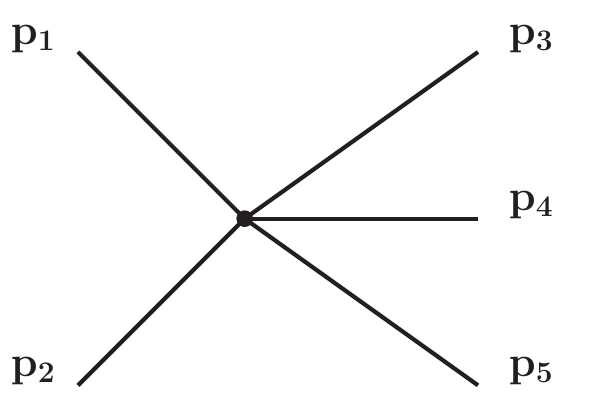}\\[8pt]
\includegraphics[width=\columnwidth]{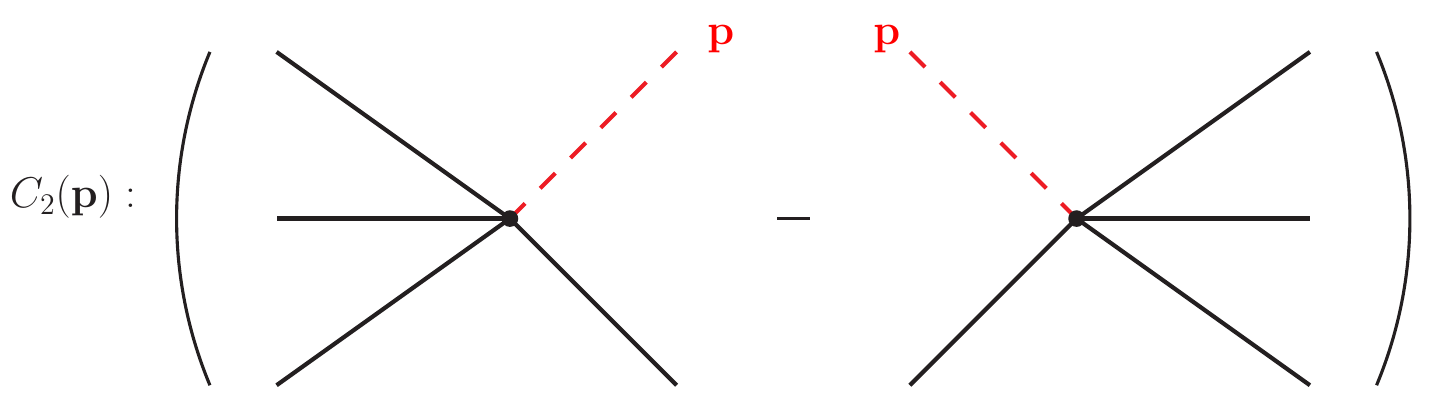}\\[6pt]
\includegraphics[width=\columnwidth]{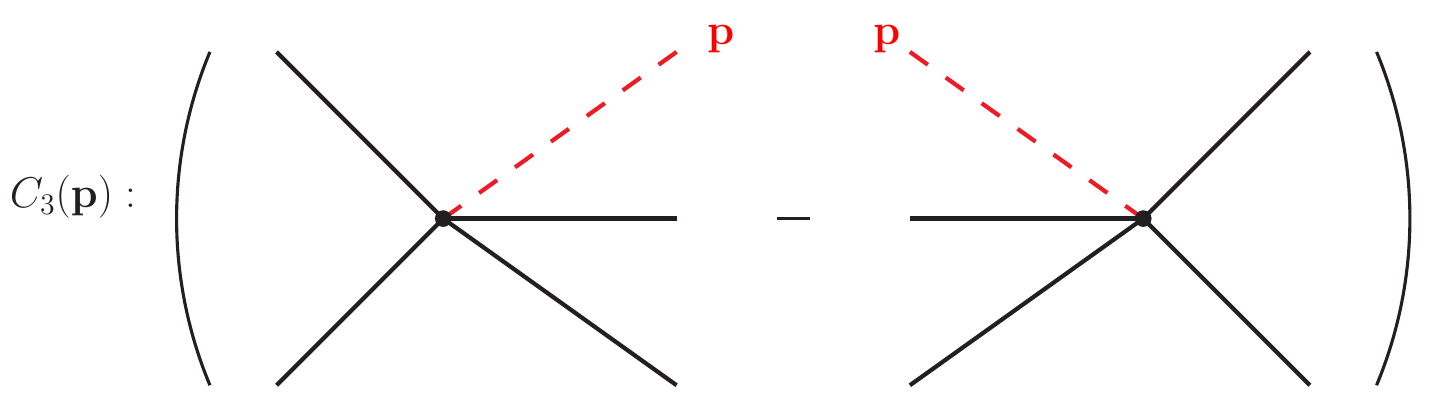}
\caption{Top: particle labeling for $\phi\phi \leftrightarrow \phi\phi\phi$. Middle and bottom: decomposition of the collision integral. The red dashed leg denotes the observed momentum $\mathbf{p}$; solid black legs are integrated over. Each term is the net rate (gain minus loss) at the observed momentum. Middle: $C_2(\mathbf{p})$, with $\mathbf{p}$ on the 2-particle side. Bottom: $C_3(\mathbf{p})$, with $\mathbf{p}$ on the 3-particle side. The full collision integral is $C_\phi(\mathbf{p}) = 2\,C_2 + 3\,C_3$.}
\label{fig:decomposition}
\end{figure}

\subsection{Identical-particle decomposition}
\label{sec:identical}

When identical particles appear in a process, the full collision integral for species $a$ at momentum $\mathbf{p}$ receives contributions from every position occupied by species $a$. For $\phi\phi \leftrightarrow \phi\phi\phi$, a scattering event affects $f_\phi(\mathbf{p})$ whenever $\mathbf{p}$ coincides with any of $\mathbf{p}_1, \ldots, \mathbf{p}_5$. The full collision integral is
\begin{equation}
\label{eq:C_full}
C_a(\mathbf{p}) = \sum_{k=1}^{n_{\rm total}} \delta_{a,s_k} \; C^{(k)}(\mathbf{p})\,,
\end{equation}
where $s_k$ denotes the species at position $k$.

By permutation symmetry of the dummy integration variables, all positions on the same side of the vertex yield the same integral. For a process with $n_\alpha$ and $n_\beta$ particles on the two sides ($n_\alpha + n_\beta = n_{\rm total}$), we define $C_{n_\alpha}(\mathbf{p})$ as $C^{(k)}(\mathbf{p})$ with the observed momentum $\mathbf{p}$ placed on the $n_\alpha$-particle side, and likewise for $C_{n_\beta}$. Note that this definition depends only on the number of legs at each side of the vertex, not on the direction of the reaction. For $\phi\phi \leftrightarrow \phi\phi\phi$, the two sides have $n_\alpha = 2$ and $n_\beta = 3$, giving $C_2(\mathbf{p})$ and $C_3(\mathbf{p})$ as illustrated in Fig.~\ref{fig:decomposition}. The full collision integral simplifies to
\begin{equation}
\label{eq:C_slot}
\boxed{C_a(\mathbf{p}) = n_\alpha^a \, C_{n_\alpha}(\mathbf{p}) + n_\beta^a \, C_{n_\beta}(\mathbf{p})\,,}
\end{equation}
where $n_\alpha^a$ ($n_\beta^a$) is the number of times species $a$ appears on the $n_\alpha$-particle ($n_\beta$-particle) side, and $|\mathcal{M}|^2$ is assumed to include the symmetry factors~\cite{Kolb:1990vq}. For the $\phi\phi \leftrightarrow \phi\phi\phi$ process this gives $C_\phi = 2\,C_2 + 3\,C_3$.

\paragraph{Why $C_2 \neq C_3$ for $n_\alpha \neq n_\beta$.}

In $C_2(\mathbf{p})$, the observed momentum $\mathbf{p}$ sits on the 2-particle side. The gain and loss terms are:
\begin{align}
\text{gain:} &\quad f_3 f_4 f_5 \cdot (1{+}f(\mathbf{p}))(1{+}f_2)\,, \\
\text{loss:} &\quad f(\mathbf{p})\,f_2 \cdot (1{+}f_3)(1{+}f_4)(1{+}f_5)\,.
\end{align}
In $C_3(\mathbf{p})$, the observed momentum sits on the 3-particle side:
\begin{align}
\text{gain:} &\quad f_1 f_2 \cdot (1{+}f(\mathbf{p}))(1{+}f_4)(1{+}f_5)\,, \\
\text{loss:} &\quad f(\mathbf{p})\,f_4 f_5 \cdot (1{+}f_1)(1{+}f_2)\,.
\end{align}
Both the role of $f(\mathbf{p})$ and the phase-space integration differ, so $C_2(\mathbf{p}) \neq C_3(\mathbf{p})$.

For \emph{symmetric} processes ($n_\alpha = n_\beta$) with identical particles on both sides, relabeling shows $C_{n_\alpha} = C_{n_\beta}$. The sum over all $n_{\rm total}$ positions then gives $n_{\rm total}$ identical contributions, but the two sides are related by exchange symmetry of the reaction, so the distinct contribution is counted with multiplicity $n_\alpha^a$:
\begin{equation}
C_a = n_\alpha^a \, C_{n_\alpha} \quad (\text{for } n_\alpha = n_\beta,\; \text{identical on both sides})\,.
\end{equation}
For $\phi\phi \leftrightarrow \phi\phi$ this gives a multiplicity of~$2$.

\paragraph{Energy conservation.} Integrating $E\,C_a(p)$ over all momenta using Eq.~\eqref{eq:C_full}:
\begin{equation}
\int \frac{d^3p}{(2\pi)^3} \, E \, C_a(p) \propto \int d\Pi_{\rm all} \left(\sum_{i \in \rm in} E_i - \sum_{j \in \rm out} E_j\right) |{\cal M}|^2 \,\Lambda\,,
\end{equation}
which vanishes by the energy-momentum delta function. This requires the complete sum over all positions; omitting $C_2$ for a $2 \leftrightarrow 3$ process gives a nonzero result.

\paragraph{Relation to the integrated equation.} This subtlety does not arise in the number density equation, where the momentum integration absorbs the sum over positions~\cite{Kolb:1990vq}. It is specific to the momentum-resolved Boltzmann equation with $n_\alpha \neq n_\beta$. Previous solvers~\cite{Ala-Mattinen:2022nuj,Binder:2021bmg,Du:2021jcj} have treated only $2 \to 2$ (and $1 \to 2$) processes where $C_{n_\alpha} = C_{n_\beta}$. Explicit collision terms for specific number-changing reactions have been given previously~\cite{Carrington:2004tm,Cervantes:2024ipg}; matching the symmetry-factor conventions, our decomposition reproduces the explicit $2 \leftrightarrow 3$ result.

\subsection{Monte Carlo evaluation of the collision integral}

For isotropic distributions $f(|\mathbf{p}|)$, we fix the observed particle along the $z$-axis: $\mathbf{p}_a = p_a\,\hat{z}$. Each remaining particle (except one determined by momentum conservation) is parametrized in spherical coordinates $(r_k, \theta_k, \phi_k)$ with Jacobian $r_k^2 \sin\theta_k$. This gives a $(3 \times (n_{\rm total} - 2))$-dimensional integral: 6 dimensions for $2 \to 2$, 9 for $2 \to 3$, and $3(n-2)$ in general.

Momentum conservation is enforced exactly by expressing the conserved particle's momentum as
\begin{equation}
\mathbf{p}_{\rm cons} = \sum_{\substack{i \in {\rm in} \\ i \neq {\rm cons}}} \mathbf{p}_i \;-\; \sum_{\substack{j \in {\rm out} \\ j \neq {\rm cons}}} \mathbf{p}_j\,,
\end{equation}
with the opposite sign for a conserved particle in the initial state. The energy of the conserved particle is then $E_{\rm cons} = \sqrt{|\mathbf{p}_{\rm cons}|^2 + m_{\rm cons}^2}$, and energy conservation is imposed via a Gaussian representation of the delta function:
\begin{equation}
\label{eq:delta}
\delta(E_{\rm in} - E_{\rm out}) \approx \frac{1}{\sigma_E \sqrt{2\pi}} \exp\!\left(-\frac{(E_{\rm in} - E_{\rm out})^2}{2\sigma_E^2}\right),
\end{equation}
where $\sigma_E$ is a width parameter with dimensions of energy. Enforcing energy conservation exactly through the delta function would remove one integration dimension, but it requires solving for one momentum magnitude, introducing case-dependent roots that do not generalize to arbitrary $n \to m$. The Gaussian keeps a uniform treatment for any multiplicity at the cost of a finite width: a smaller $\sigma_E$ enforces energy conservation more strictly but confines the integrand to a thinner shell, which is harder to sample. The balance between the two is discussed in Appendix~\ref{app:numerics}. The choice of which particle to assign as the conserved particle affects the efficiency of the Monte Carlo integration. For asymmetric processes ($n_{\rm in} \neq n_{\rm out}$), the conserved particle is placed on the side with fewer particles (e.g.\ the initial state for $2 \to 3$ processes). This ensures that the conserved particle's momentum, determined as the difference between the total momenta of the two sides, is kinematically allowed over a broad region of the integrated phase space.

The gain and loss contributions are integrated with \emph{separate} \textsc{Vegas} integrators, following the strategy of Ref.~\cite{Ala-Mattinen:2022nuj}. This avoids numerical cancellation between two large terms and allows each integrator to adapt its importance sampling map independently.

The phase-space factor for the full integral, after enforcing momentum conservation, is
\begin{equation}
\frac{\prod_{k} r_k^2 \sin\theta_k}{(2\pi)^{3(n_{\rm total}-1)-4} \prod_{i=1}^{n_{\rm total}} 2E_i}\,,
\end{equation}
where the product over $k$ runs over the integrated particles and the $(2\pi)$ power follows from counting the measure factors.

\section{Implementation}
\label{sec:implementation}

\subsection{Code structure}

\textsc{Best} is implemented in Python, using \texttt{numpy} for vectorized array operations, \texttt{scipy} for interpolation and quadrature, \texttt{mpi4py} for parallelization, and the \texttt{vegas} library~\cite{Lepage:2020tgj} for adaptive Monte Carlo integration. The entire solver is contained in a single file with one main class, requiring no installation beyond the dependencies listed above. The code is intentionally kept transparent and minimal: users are encouraged to read and modify the source directly to extract additional observables or adapt the solver to their specific needs.

Species are initialized on a momentum grid (logarithmic or linear) spanning $[q_{\rm min}, q_{\rm max}]$ with $N_{\rm grid}$ points. Each species carries a quantum statistics label (boson, fermion, or Maxwell-Boltzmann), a mass (possibly time-dependent via a user-supplied function), and its phase-space distribution $f(q)$.

Processes are registered by specifying the initial and final species, the squared matrix element $|M|^2$ as a callable function of the momenta and coupling, and integration parameters (\texttt{neval}, \texttt{nitn}, \texttt{alpha}, \texttt{delta\_width}). The matrix element is expected to include all symmetry factors for identical particles. The code automatically determines the integration dimensionality from the number of particles, and detects whether the identical-particle decomposition of Eq.~\eqref{eq:C_slot} is needed by comparing the species multiplicities on each side.

\subsection{Interpolation and extrapolation}
\label{sec:interp}

Between grid points, the distribution function is accessed through a cubic spline interpolator in linear scale. Outside the grid, extrapolation is performed by fitting the form
\begin{equation}
\label{eq:extrap}
\log\!\left(\frac{1}{f} + \eta\right) = a + b\,E(p)\,,
\end{equation}
where $\eta = +1$ for bosons, $\eta = -1$ for fermions, and $\eta = 0$ for Maxwell-Boltzmann statistics, and $E(p) = \sqrt{p^2 + m^2}$. The parameters $(a,b)$ are determined by linear fits to the low-momentum and high-momentum ends of the grid separately. This form is exact for equilibrium distributions ($f_{\rm BE} = 1/(e^{E/T}-1)$ gives $\log(1/f+1) = E/T$) and provides physically motivated extrapolation for near-equilibrium distributions.

\subsection{Time integration}

The distribution is evolved in log-space to ensure positivity. For the Euler method:
\begin{equation}
\log f(t+\Delta t) = \log f(t) + \Delta t \, \frac{C[f]}{f}\,.
\end{equation}
The default method is Heun's method (second-order), which uses an Euler predictor step to compute $f^*$, then evaluates the collision integral at $f^*$ to obtain $k_2$, and combines:
\begin{equation}
\label{eq:heun}
\log f(t+\Delta t) = \log f(t) + \frac{\Delta t}{2}\left(\frac{k_1}{f} + \frac{k_2}{f^*}\right),
\end{equation}
where $k_1 = C[f(t)]$ and $k_2 = C[f^*]$. The time step is adapted to satisfy
\begin{equation}
\max_p \left|\Delta t\,\frac{C[f](p)}{f(p)}\right| < \varepsilon\,,
\end{equation}
where $\varepsilon$ limits the fractional change of $f$ per step, set to $0.3$ in this work, with smaller values improving accuracy at the cost of more steps.

\begin{figure*}[t]
\centering
\includegraphics[width=0.48\textwidth]{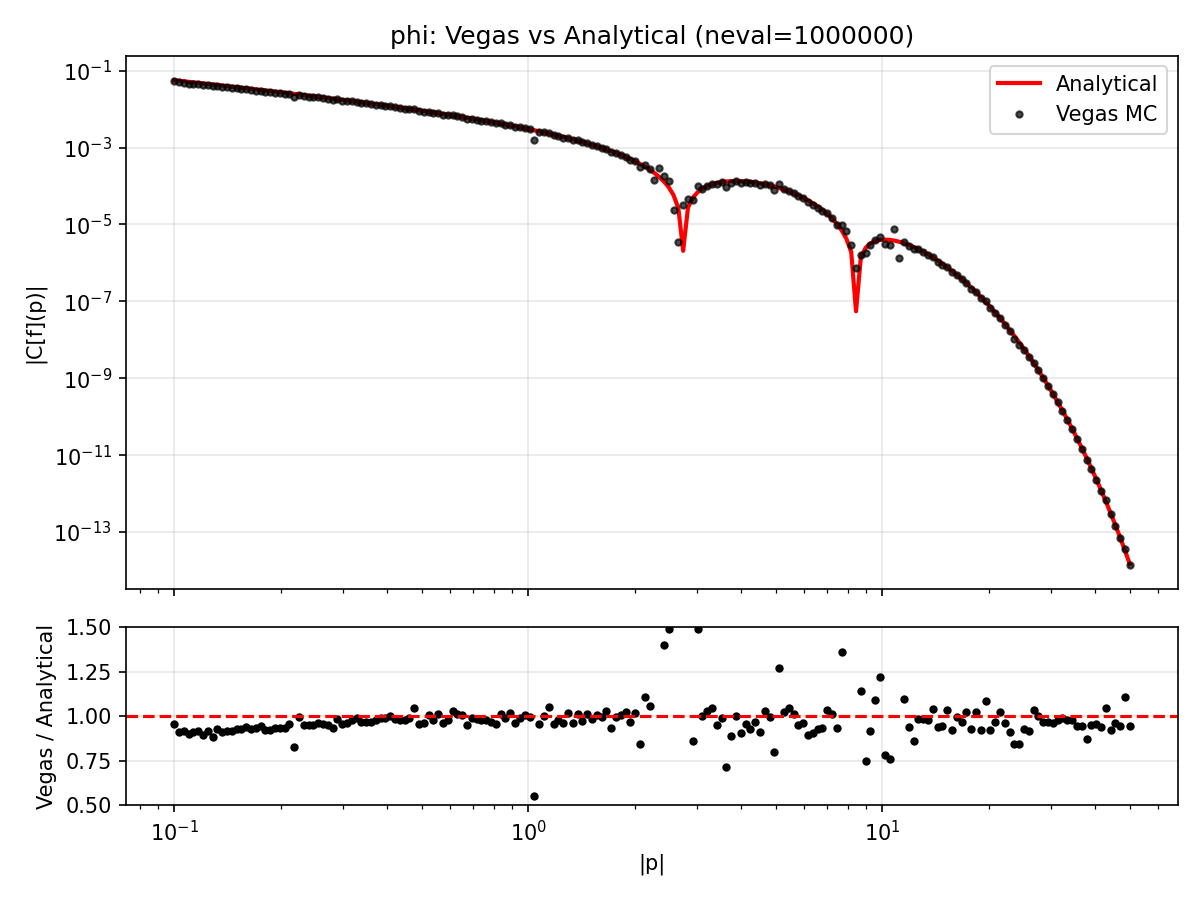}
\includegraphics[width=0.48\textwidth]{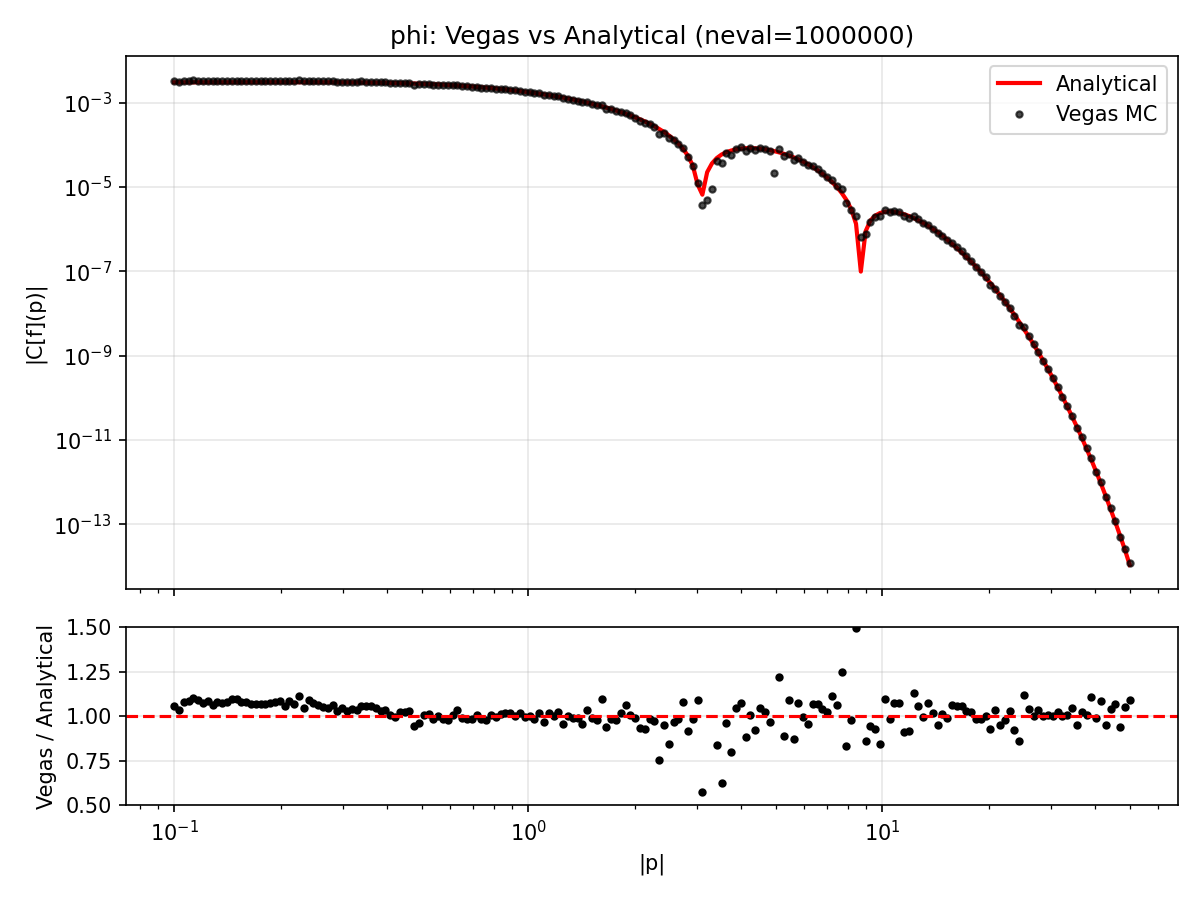}
\caption{Comparison of the collision rate $|C[f](p)|$ computed by \textsc{Vegas} Monte Carlo (points) and the semi-analytical method (solid line), for a non-thermal initial distribution with $\lambda = 1$ and $H = 0$. Left panels: massless ($m_\phi = 0$). Right panels: massive ($m_\phi = 1$). Upper panels show the absolute rate on a log-log scale; lower panels show the ratio. Both methods include the identical-particle multiplicity factor of~2. For both cases, agreement is within a few percent where the collision rate is large; increased scatter appears near the zero crossings of $C[f]$ where the Monte Carlo signal-to-noise ratio is low.}
\label{fig:benchmark}
\end{figure*}

\subsection{Parallelization}
\label{sec:parallel}

We parallelize over momentum grid points. The MPI world communicator is split into $N_r$ groups, and each group evaluates the collision integral at its assigned grid points. The \textsc{Vegas} \texttt{mpi=True} option remains enabled within each sub-communicator, so that groups with more than one rank can still distribute integrand evaluations internally. After all groups complete, results are collected via \texttt{MPI\_Allreduce}. For $N_r$ grid points and $N_{\rm proc}$ MPI ranks, the wall time scales as
\begin{equation}
T \approx \frac{N_r}{N_{\rm proc}} \times t_{\rm eval}\,,
\end{equation}
where $t_{\rm eval}$ is the time per grid point, which depends on $N_{\rm eval}$ and the integration dimensionality. A key optimization is \emph{integrator reuse}: the \textsc{Vegas} adaptive map is preserved across grid points and time steps. Since the integrand structure varies slowly along the momentum grid and between successive time steps, the importance sampling distribution remains a good approximation, requiring only minor refinement (controlled by the \texttt{alpha} parameter).

\subsection{Cosmological expansion}
\label{sec:expansion}

The code supports cosmological expansion through a user-defined scale factor $a(t)$. A built-in option sets radiation domination, $a(t) = a_0 (t/t_0)^{1/2}$. The physical momenta in the collision integral are computed as $p = q/a(t)$, where $q$ is the comoving momentum stored on the grid. The phase-space Jacobian includes the appropriate factor of $a^{-3}$ per integrated particle to account for the transformation from comoving to physical momentum volume: $d^3p = d^3q/a^3$. The comoving momentum of the conserved particle is recovered as $q_{\rm cons} = |\mathbf{p}_{\rm cons}|\times a$.

\subsection{Output and checkpointing}

The solver writes a single Python pickle file that serves both as a checkpoint for restart and as the primary output. It contains the distribution functions, momentum grids, species and process definitions, \textsc{Vegas} integrator states, and the full evolution history; a plotting script included with the code illustrates how to read it and reproduce evolution plots. Matrix element functions are restored by name lookup in the caller's namespace.

\section{Semi-analytical benchmark}
\label{sec:analytical}

For $2 \to 2$ processes with general masses and constant $|M|^2$, the collision integral can be reduced to two dimensions following the methods developed for neutrino transport~\cite{Yueh:1976,Hannestad:1995rs} and applied to dark matter in Ref.~\cite{Ala-Mattinen:2022nuj}. The backward term becomes
\begin{multline}
\label{eq:CBW}
C_{\rm BW}(p_1) = \frac{2}{(2\pi)^4} \frac{1}{2E_1} \int dp_3 \frac{p_3^2}{2E_3} \int dp_4 \frac{p_4^2}{2E_4} \\
\times F(p_1,p_3,p_4)\, f_3 f_4 (1{\pm}f_1)(1{\pm}f_2)\,,
\end{multline}
where energy conservation fixes $E_2 = E_3 + E_4 - E_1$ exactly (with $p_2 = \sqrt{E_2^2 - m_2^2}$), and the angular-integrated kinematic function is
\begin{equation}
F(p_1,p_3,p_4) = \int_{-1}^{1}\!\! d\!\cos\theta\;\frac{|M|^2 \pi}{\sqrt{-a(\cos\theta)}}\;\Theta(b^2-4ac)\,,
\end{equation}
with $a$, $b$, $c$ quadratic forms in the kinematic variables as defined in Ref.~\cite{Ala-Mattinen:2022nuj}. The forward term has an analogous form with integration variables $(p_2, p_3)$.

This semi-analytical method enforces energy conservation exactly (no Gaussian approximation), providing an independent check on the Monte Carlo integration accuracy and the systematic error from the delta function width. We precompute the kinematic function $F$ on an $n_F \times n_F$ grid and cache it, since $F$ depends only on the momenta and $|M|^2$, not on the distribution function. The remaining 2D integral over the momentum magnitudes is evaluated using the trapezoidal rule on a logarithmic grid.

This benchmark computes a single-position contribution $C^{(k)}$. For $\phi\phi \leftrightarrow \phi\phi$ with identical particles, the full rate is obtained by multiplying by the multiplicity $n_\alpha^a = 2$, consistent with Eq.~\eqref{eq:C_slot} and the exchange symmetry $C_{n_\alpha} = C_{n_\beta}$.

\begin{figure*}[t]
\centering
\includegraphics[width=\textwidth]{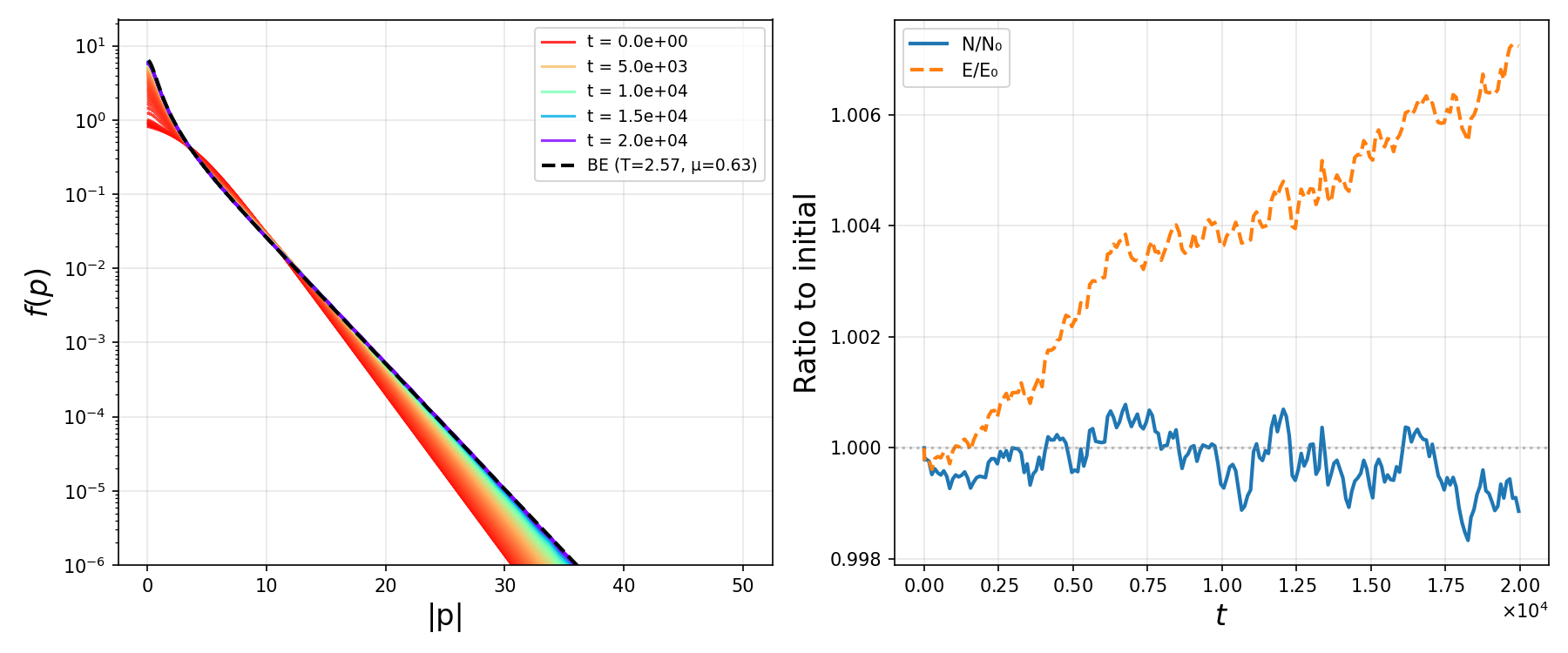}
\caption{Thermalization via $2 \to 2$ elastic scattering with massive $\phi$ ($m_\phi = 1$, $\lambda = 1$, $H = 0$). Left panel: evolution of $f(p)$ from a non-thermal initial condition (red) toward the Bose-Einstein distribution (black dashed), with $T_{\rm eq}$ and $\mu$ determined by the conserved energy and particle number. Right panel: conservation of particle number $N/N_0$ (solid) and energy $E/E_0$ (dashed).}
\label{fig:therm_2to2}
\end{figure*}

\section{Results}
\label{sec:results}

\subsection{Setup}

We consider a single scalar species $\phi$ in flat spacetime ($H = 0$) to isolate the collision dynamics from cosmological expansion. For the $2 \to 2$ benchmark comparison, we validate both the massless ($m_\phi = 0$) and massive ($m_\phi = 1$) cases with quartic self-interaction $\phi\phi \to \phi\phi$ and constant matrix element $|M|^2 = \lambda^2$ with $\lambda = 1$. For the thermalization studies and the $2 \leftrightarrow 3$ demonstration, we use massive particles ($m_\phi = 1$) with a five-point contact interaction $\phi\phi \leftrightarrow \phi\phi\phi$ and constant $|M|^2 = \lambda_5^2$ with $\lambda_5 = 1$. All runs start from a non-thermal initial distribution $f_0(p) = 1/(1 + e^{(p-p_0)/w})$ with $p_0 = 3$, $w = 2$. The comoving momentum grid spans $q \in [0.1, 50]$ with $N_{\rm grid} = 272$ logarithmically spaced points. With $H = 0$, the thermalization timescale is set entirely by the coupling strength; in a cosmological setting, thermalization requires the interaction rate $\Gamma \sim n \langle\sigma v\rangle$ to exceed the Hubble rate $H$.

\subsection{Vegas vs.\ analytical: $2 \to 2$ benchmark}

Figure~\ref{fig:benchmark} compares the collision rate computed by the \textsc{Vegas} Monte Carlo and the semi-analytical method described in Sec.~\ref{sec:analytical}. The two methods agree to within a few percent where the rate is sizable. Increased scatter appears near the zero crossings of the collision rate, where the signal-to-noise ratio of the Monte Carlo estimator is low.

\subsection{Thermalization dynamics: $2 \to 2$}

Starting from a non-thermal initial condition, the distribution evolves toward the Bose-Einstein form as shown in Fig.~\ref{fig:therm_2to2}. The distribution converges to $f_{\rm BE}(p) = 1/(e^{(E-\mu)/T_{\rm eq}}-1)$, where the equilibrium temperature $T_{\rm eq}$ and chemical potential $\mu$ are determined by the conserved total energy and particle number densities. The low-momentum region is enhanced by Bose stimulation, while the high-momentum tail is depleted. Particle number and energy are conserved to within $\sim 1\%$ throughout the evolution.

\subsection{$2 \leftrightarrow 3$ number-changing process}
\label{sec:2to3}

The $\phi\phi \leftrightarrow \phi\phi\phi$ process provides the key test of the identical-particle decomposition derived in Sec.~\ref{sec:identical}. Species $\phi$ appears $n_\alpha^a = 2$ times on the 2-particle side and $n_\beta^a = 3$ times on the 3-particle side, so Eq.~\eqref{eq:C_slot} gives $C_\phi = 2\,C_2 + 3\,C_3$, as illustrated in Fig.~\ref{fig:decomposition}.

\begin{figure*}[t]
\centering
\includegraphics[width=\textwidth]{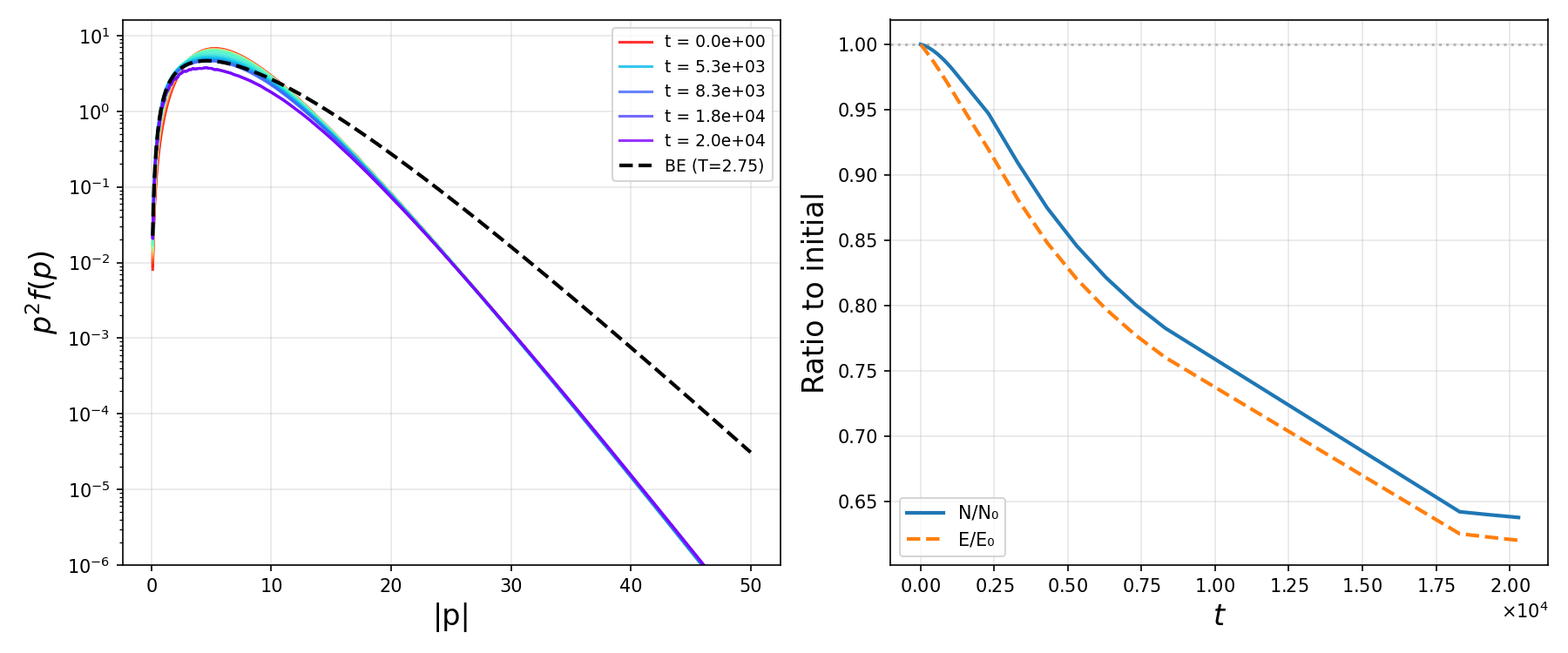}
\includegraphics[width=\textwidth]{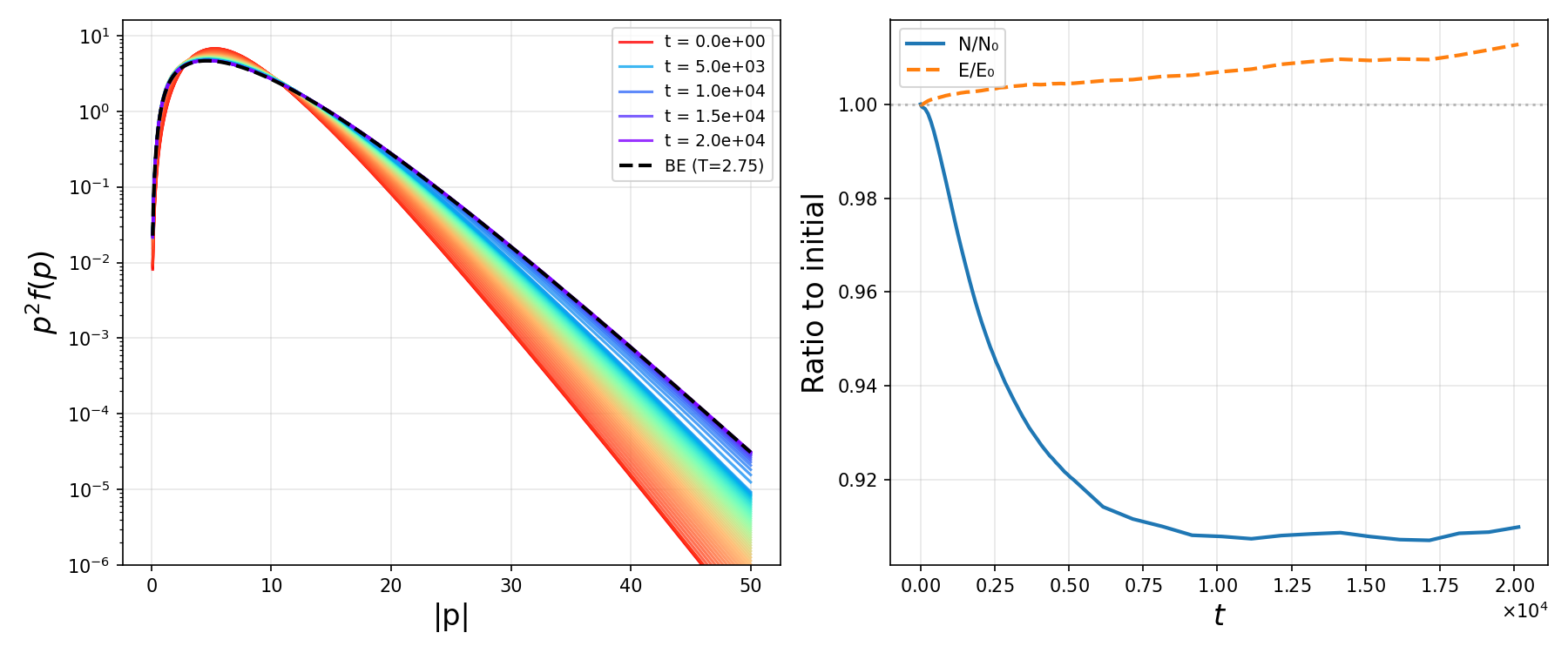}
\caption{$2 \leftrightarrow 3$ number-changing process with massive $\phi$ ($m_\phi = 1$, $\lambda_5 = 1$, $H = 0$). Left panels: evolution of $p^2 f(p)$ from a non-thermal initial condition (red) toward the equilibrium Bose-Einstein distribution with $\mu = 0$ (black dashed). Right panels: conservation of particle number $N/N_0$ (solid) and energy $E/E_0$ (dashed). Top: \emph{incorrect} calculation using only $C_3$ (shown for illustration), leading to $\sim 40\%$ energy loss. Bottom: correct decomposition $2\,C_2 + 3\,C_3$, which restores energy conservation.}
\label{fig:2to3}
\end{figure*}

Figure~\ref{fig:2to3} demonstrates the necessity of the full decomposition. Using only $C_3$ (top panels)---as one might naively implement---leads to $\sim 40\%$ energy loss over the simulation. This is not a Monte Carlo artifact but a consequence of the missing $C_2$ contributions: the energy integral $\int E\,C_3\,p^2\,dp$ does not vanish because it does not include the complete sum over all positions. Including the correct combination $2\,C_2 + 3\,C_3$ (bottom panels) restores energy conservation to within $\sim 1\%$, consistent with Monte Carlo statistical noise.

The bottom-left panel shows the resulting thermalization. The distribution approaches a Bose-Einstein form with the temperature set by energy conservation and the chemical potential driven toward $\mu = 0$ by the number-changing process, since chemical equilibrium for $\phi\phi \leftrightarrow \phi\phi\phi$ requires $3\mu = 2\mu$, hence $\mu = 0$~\cite{Profumo:2025uvx}. The particle number decreases over time as the $3 \to 2$ direction is favored for the chosen initial condition. In the top panels, the incorrect calculation not only violates energy conservation but also fails to reach the correct equilibrium, as the missing $C_2$ contributions distort the balance between production and absorption.

\subsection{Computational performance}
\label{sec:performance}
The parallelization strategy described in Sec.~\ref{sec:parallel} yields near-linear scaling up to $N_{\rm proc} \approx N_{\rm grid}$, as each momentum grid point is independent. For the $2 \to 2$ benchmarks ($N_{\rm grid} = 272$, $N_{\rm eval} = 10^6$, 6 dimensions), a single Heun step takes $\sim 50$\,s with 272 MPI ranks. For the $2 \to 3$ process ($N_{\rm grid} = 136$, $N_{\rm eval} = 10^7$, 9 dimensions), a single Euler step takes $\sim 300$\,s with 544 MPI ranks, dominated by the two separate passes required for $C_2$ and $C_3$. The thermalization runs shown in Figs.~\ref{fig:therm_2to2} and \ref{fig:2to3} were performed on the KISTI Nurion supercomputer, each completing in a few hours of total wall time.

\section{Discussion}
\label{sec:discussion}

We have presented \textsc{Best}\footnote{The collision integral for general $n \to m$ processes is a high-dimensional, noisy integral that cannot be simplified analytically beyond $2 \to 2$. The code makes the \textit{best} of what is computationally feasible, combining adaptive Monte Carlo integration, massively parallel evaluation, and integrator reuse to push the numerical evaluation of thermalization dynamics as far as current methods allow.}, a general-purpose solver for momentum-resolved Boltzmann equations that handles arbitrary $n \to m$ processes via direct Monte Carlo evaluation of the collision integral.

A key technical ingredient is the identical-particle decomposition, Eq.~\eqref{eq:C_slot}, which is essential for processes with $n_\alpha \neq n_\beta$. This requirement is a direct consequence of the Boltzmann equation. For the $2 \leftrightarrow 3$ cannibal process studied here, neglecting this decomposition leads to $\sim 40\%$ energy nonconservation, demonstrating that it is not a small correction but a qualitative requirement for the correct collision integral.

The key advantages of the code over existing solvers are:

\begin{enumerate}
\item \emph{Generality}: The code treats $2 \to 2$, $2 \to 3$, $3 \to 2$, and higher-multiplicity processes on equal footing, with the integration dimensionality determined automatically.
\item \emph{Correct identical-particle treatment}: The decomposition of Eq.~\eqref{eq:C_slot} is applied automatically when species multiplicities differ between the two sides of the reaction.
\item \emph{Full quantum statistics}: Bose enhancement and Pauli blocking are included without approximation in both gain and loss terms.
\item \emph{Massive particles}: The code supports arbitrary masses, including time-dependent masses relevant for phase transitions and thermal corrections.
\item \emph{Multi-species}: Multiple coupled species can be evolved simultaneously. Conversions, decays, and inelastic scatterings use the same interface, differing only in the assigned input and output species; the cost scales linearly with the number of processes, while the dominant cost is set by the phase-space dimensionality of each.
\end{enumerate}

The primary numerical challenge is the Monte Carlo evaluation of the collision integral in high dimensions. For processes with many final-state particles, the energy-conserving region of phase space occupies a small fraction of the integration domain, and the \textsc{Vegas} adaptive map requires sufficient evaluations to locate it efficiently. The placement of the conserved particle on the side with fewer particles (Sec.~\ref{sec:formalism}) is crucial for maintaining a broad physical phase space. Similarly, the extrapolation of the distribution function outside the momentum grid must be handled with care, as the collision integral is sensitive to the tail behavior of $f(p)$ through the statistical factors.

Near the zero crossings of the collision rate, the Monte Carlo signal-to-noise ratio degrades, as seen in Fig.~\ref{fig:benchmark}. For $2 \to 2$ processes, the semi-analytical method with exact energy conservation provides a useful cross-check.

For massless bosons, the equilibrium distribution $f_{\rm BE}(p) = T/p$ diverges at low momenta, which poses a challenge for the Monte Carlo evaluation: large occupation numbers amplify the Bose enhancement factors in the collision integral, increasing the variance of the estimator. In physical applications, this infrared divergence is regulated by thermal mass corrections $m_{\rm th}^2 \sim \lambda T^2$~\cite{Lebedev:2021xey}, which arise from resummed self-energy diagrams and ensure a finite occupation number at zero momentum. The code supports thermal masses through the time-dependent mass interface, allowing the user to supply the appropriate correction for a given model. For the results presented here, we use massive particles ($m_\phi = 1$) throughout the thermalization studies, for which the equilibrium distribution remains finite at all momenta.

The results presented here use flat spacetime ($H = 0$) to isolate the collision dynamics. In a cosmological setting, thermalization requires the interaction rate $\Gamma \sim n\langle\sigma v\rangle$ to exceed the Hubble rate $H \sim 1/(2t)$ in radiation domination. When $\Gamma < H$, the expansion redshifts momenta faster than collisions can redistribute them, and the distribution freezes out. The code supports radiation-dominated expansion and can track this competition explicitly; a study of the $\Gamma/H$ dependence and the resulting freeze-out spectrum is left for future work.

The framework is designed to be extended to a range of cosmological applications where momentum-resolved evolution beyond $2 \to 2$ is required. These include freeze-in production through multi-body final states, cannibal dark matter with $3 \to 2$ number-changing interactions~\cite{Carlson:1992fn,Hochberg:2014dra}, strongly interacting massive particle (SIMP) scenarios~\cite{Hochberg:2014kqa}, semi-annihilation processes~\cite{DEramo:2010keq}, and scenarios where kinetic and chemical equilibrium are lost simultaneously~\cite{Duch:2017nbe,Binder:2017rgn,Profumo:2025uvx}. Such processes are increasingly studied at the phase-space level. The support for time-dependent masses further enables the study of thermalization across cosmological phase transitions, where the spectrum is shaped by a changing dispersion relation during the transition itself. A further application is reheating after inflation, where decay products (e.g.\ of the inflaton) thermalize from an out-of-equilibrium state, refining the simplified estimate of the reheating temperature based on instantaneous thermalization.

The code is publicly available at \url{https://github.com/best-hep/best}.

\begin{acknowledgments}
The author thanks Oleg Lebedev and Jong-Chul Park for their guidance and support of this project.
This work was supported by the National Research Foundation of Korea (NRF) grant funded by the Korea government (MSIT) (RS-2024-00356960, RS-2025-00559197) and by the National Supercomputing Center with supercomputing resources including technical support (KSC-2025-CRE-0282).
The use of Claude (Anthropic) for assistance in code development and debugging is acknowledged.
\end{acknowledgments}

\appendix

\section{Phase-space construction for general $n \to m$ processes}
\label{app:phasespace}

\subsection{Coordinate system and integration variables}

We fix the observed particle's momentum along the $z$-axis: $\mathbf{p}_{\rm obs} = p_{\rm obs}\,\hat{z}$. Each integrated particle $k$ is parametrized in spherical coordinates:
\begin{align}
p_{k,x} &= r_k \sin\theta_k \cos\phi_k\,, \\
p_{k,y} &= r_k \sin\theta_k \sin\phi_k\,, \\
p_{k,z} &= r_k \cos\theta_k\,,
\end{align}
with $r_k \in [q_{\rm min}, q_{\rm max}]$, $\theta_k \in [0, \pi]$, $\phi_k \in [0, 2\pi)$, and Jacobian $r_k^2 \sin\theta_k$.

\subsection{Particle roles}

For a process with $n_{\rm total} = n_{\rm in} + n_{\rm out}$ particles, we label all particles with indices $i = 0, \ldots, n_{\rm total}-1$, where $i < n_{\rm in}$ are initial-state and $i \geq n_{\rm in}$ are final-state particles. Three roles are assigned:

\textit{Observed particle} (index $i_{\rm obs}$): Fixed at the momentum $\mathbf{p}$ where we evaluate $C^{(k)}[f]$. When computing $C_{n_{\rm in}}$, the observed particle is placed on the $n_{\rm in}$-particle side; when computing $C_{n_{\rm out}}$, on the $n_{\rm out}$-particle side.

\textit{Conserved particle} (index $i_{\rm cons}$): Determined by 3-momentum conservation. For asymmetric processes ($n_{\rm in} \neq n_{\rm out}$), it is placed on the side with fewer particles. For symmetric processes ($n_{\rm in} = n_{\rm out}$), it is placed on the same side as the observed particle. If the observed particle occupies the only available position on the fewer-particle side, the conserved particle is placed on the opposite side as a fallback. This placement ensures that the conserved particle's momentum, reconstructed from the other particles via momentum conservation, generically yields a physical (positive, on-shell) energy over a broad region of the integrated phase space.

\textit{Integrated particles}: All remaining $n_{\rm total} - 2$ particles, sampled by \textsc{Vegas} in spherical coordinates. This gives a $3(n_{\rm total}-2)$-dimensional integral.

\subsection{Momentum conservation}

The conserved particle's 3-momentum is
\begin{equation}
\mathbf{p}_{\rm cons} = \begin{cases}
\displaystyle\sum_{\substack{j \in {\rm out} \\ j \neq {\rm cons}}} \mathbf{p}_j - \sum_{\substack{i \in {\rm in} \\ i \neq {\rm cons}}} \mathbf{p}_i & \text{if } i_{\rm cons} < n_{\rm in}\,, \\[12pt]
\displaystyle\sum_{\substack{i \in {\rm in} \\ i \neq {\rm cons}}} \mathbf{p}_i - \sum_{\substack{j \in {\rm out} \\ j \neq {\rm cons}}} \mathbf{p}_j & \text{if } i_{\rm cons} \geq n_{\rm in}\,,
\end{cases}
\end{equation}
Its comoving momentum is $q_{\rm cons} = |\mathbf{p}_{\rm cons}| \times a(t)$, and its energy is $E_{\rm cons} = \sqrt{|\mathbf{p}_{\rm cons}|^2 + m_{\rm cons}^2}$.

\subsection{Energy conservation}

Energy conservation is approximated via Eq.~\eqref{eq:delta}. The normalization ensures that the Gaussian integrates to unity in the limit $\sigma_E \to 0$, recovering the exact delta function. Points with $E_k < E_{\rm min}^{\rm cut}$ for any particle are discarded.

\subsection{Phase-space factor}

Combining the momentum-space measures, the phase-space weight per Monte Carlo point is
\begin{equation}
w = \frac{\prod_{k \in \rm integrated} r_k^2 \sin\theta_k}{(2\pi)^{3(n_{\rm total}-1)-4}\;\prod_{i=1}^{n_{\rm total}} 2E_i} \times \delta_{\rm Gauss}(E_{\rm in} - E_{\rm out})\,.
\end{equation}

\section{Semi-analytical $2 \to 2$ collision integral}
\label{app:analytical}

For particles with general masses and isotropic distributions, the $2 \to 2$ collision integral can be reduced from 9 to 2 dimensions following the approach of Refs.~\cite{Yueh:1976,Hannestad:1995rs,Ala-Mattinen:2022nuj}. We outline the key steps and establish notation.

\subsection{Backward term}

Starting from Eq.~\eqref{eq:collision_single} for the backward (gain) contribution, we fix $p_1$ and integrate over $d^3p_3\,d^3p_4$, using the four-momentum delta function to eliminate $d^3p_2$. This leaves a constraint $\delta(p_2^2 - m_2^2)$ which enforces $E_2 = E_3 + E_4 - E_1$ and $p_2 = \sqrt{E_2^2 - m_2^2}$. Choosing $\hat{p}_1$ as the polar axis and parametrizing $\mathbf{p}_3$ and $\mathbf{p}_4$ as in Ref.~\cite{Ala-Mattinen:2022nuj}, the integral over the azimuthal angle $\beta$ of $\mathbf{p}_4$ is performed using the delta function. The remaining integral over $\cos\alpha$ (the polar angle of $\mathbf{p}_4$ relative to $\hat{p}_1$) is bounded by $b^2 - 4ac \geq 0$, yielding the kinematic function
\begin{equation}
F(p_1, p_3, p_4) = \int_{-1}^{1} d\cos\theta \; \frac{|M|^2\,\pi}{\sqrt{-a}} \; \Theta(b^2 - 4ac)\,,
\end{equation}
where
\begin{align}
Q &= m_1^2 - m_2^2 + m_3^2 + m_4^2\,, \\
\gamma &= E_3 E_4 - E_1 E_3 - E_1 E_4\,, \\
\epsilon &= p_1 p_3 \cos\theta\,, \quad \kappa = p_1^2 + p_3^2\,, \\
a &= p_4^2(-4\kappa + 8\epsilon)\,, \\
b &= p_4(-p_1 + \epsilon/p_1)(8\gamma + 4Q + 8\epsilon)\,, \\
c &= 4p_3^2 p_4^2 \sin^2\!\theta - (2(\gamma + \epsilon) + Q)^2\,.
\end{align}

\subsection{Forward term}

The forward term is obtained analogously with integration variables $(p_2, p_3)$ and energy conservation $E_4 = E_1 + E_2 - E_3$. The kinematic function $F'$ has the same functional form with replacements
\begin{align}
Q' &= m_1^2 + m_2^2 + m_3^2 - m_4^2\,, \\
\gamma' &= E_1 E_2 - E_1 E_3 - E_2 E_3\,, \\
\kappa' &= p_1^2 + p_3^2\,,
\end{align}
and $p_2$ replacing $p_4$ in the coefficients $a'$, $b'$, $c'$.

\subsection{Caching}

The kinematic function $F$ depends only on the momenta and $|M|^2$, not on the distribution function. We precompute $F$ on an $n_F \times n_F$ grid for each value of $p_1$ and cache the result. Since the grid is fixed throughout the evolution, this computation is performed once and reused at every time step. The remaining 2D integral is evaluated by the trapezoidal rule.

\section{Numerical parameters and stability}
\label{app:numerics}

The main parameters controlling the Monte Carlo evaluation are the number of evaluations per iteration (we use $10^6$ for $2 \to 2$ and $10^7$ for $2 \to 3$ processes), the number of \textsc{Vegas} iterations (we use two, which suffices with integrator reuse), the adaptation rate (we use $0.5$), and the fractional width of the Gaussian that enforces energy conservation, $\sigma_E = w\,(E_{\rm in} + E_{\rm out})/2$, with $w$ adapted during the integration as described below.

Two settings control the trade-off between accuracy and energy conservation: the number of Monte Carlo evaluations and the width of the Gaussian. Each has a clear best direction on its own. More evaluations give better accuracy, and a narrower width conserves energy better. These are not independent, however: making the width narrower also makes the integral harder to evaluate, so a narrow width needs more evaluations to stay accurate. The code handles this automatically, narrowing the width when the relative Monte Carlo error falls below a lower threshold and widening it when the error exceeds an upper threshold. As a result, more evaluations permit a narrower width and better energy conservation; with too few, the width must stay wide and energy conservation breaks down. For the $2 \to 2$ benchmark, on the order of $10^5$ evaluations with an initial width of order $10^{-2}$ gives stable evolution, while an order of magnitude fewer evaluations, or a wider initial width, is insufficient.

When the number of evaluations cannot be increased but the evolution is too slow, raising the lower threshold lets the width narrow at the accuracy that is achievable, recovering the evolution at the cost of looser energy conservation; it can be lowered again once more evaluations are afforded.

A laptop is suitable for debugging and small-scale validation with reduced parameters; production runs and timing are described in Sec.~\ref{sec:performance}.
\section{Quick start}
\label{app:quickstart}

We illustrate the basic usage of \textsc{Best} with two minimal examples.

\subsection{$2 \to 2$ elastic scattering}

\begin{lstlisting}[language=Python]
import numpy as np
from besthep import BEST

def matrix_element(momenta, coupling):
    """Constant |M|^2."""
    return np.full(momenta.shape[2],
                   coupling**2)

def init_f(r):
    """Non-thermal initial condition."""
    return 1.0 / (1 + np.exp((r - 3) / 2.0))

solver = BEST(q_min=0.1, q_max=20.0,
              n_grid=64)

solver.initialize_species(
    'phi', init_f,
    stat='boson', mass=1.0)

solver.add_process(
    'elastic',
    ['phi', 'phi'], ['phi', 'phi'],
    matrix_element,
    coupling=1.0, neval=int(1e6),
    delta_width=0.01)

for step in range(100):
    solver.evolve_step(dt=1.0, method='heun')
\end{lstlisting}

\noindent {A complete version of this example, including
checkpointing and diagnostic output, is provided as
\texttt{examples/2to2m1.py}.} Run with MPI:
\begin{lstlisting}[language=bash]
mpirun -np 4 python3 examples/2to2m1.py
\end{lstlisting}

\subsection{$2 \to 3$ number-changing process}

The only change is the process registration:
\begin{lstlisting}[language=Python]
solver.add_process(
    'cannibal',
    ['phi', 'phi'], ['phi', 'phi', 'phi'],
    matrix_element,
    coupling=1.0, neval=int(1e7),
    delta_width=0.01)
\end{lstlisting}

\noindent The integration dimensionality ($6 \to 9$), phase-space construction, and identical-particle decomposition ($C = 2\,C_2 + 3\,C_3$) are all handled automatically. For $2 \to 3$, higher \texttt{neval} is recommended due to the increased dimensionality.

\subsection{Adding cosmological expansion}

\begin{lstlisting}[language=Python]
solver.current_time = 100.0
solver.set_radiation_dominated(
    a0=1.0, t0=solver.current_time)
\end{lstlisting}

\noindent The grid stores comoving momenta $q = ap$; the collision integral uses physical momenta $p = q/a(t)$ internally.

\subsection{Multiple species and massive particles}

\begin{lstlisting}[language=Python]
solver.initialize_species(
    'chi', init_chi,
    stat='fermion', mass=5.0)
solver.initialize_species(
    'phi', init_phi,
    stat='boson', mass=1.0)

solver.add_process(
    'annihilation',
    ['chi', 'chi'], ['phi', 'phi'],
    matrix_element_ann,
    coupling=0.1, neval=int(1e6))
\end{lstlisting}

\noindent Time-dependent masses (e.g.\ across a phase transition) are set via:
\begin{lstlisting}[language=Python]
solver.set_mass_func('phi',
    lambda t: 1.0 if t > 20 else 0.0)
\end{lstlisting}

\subsection{Checkpointing}

\begin{lstlisting}[language=Python]
# Save
solver.save_checkpoint('checkpoint.pkl',
                       history=history)
# Resume
history = solver.load_checkpoint(
    'checkpoint.pkl',
    matrix_elements={
        'matrix_element': matrix_element})
\end{lstlisting}


\end{document}